\documentclass[aps,prl,showpacs,showkeys,reprint,floatfix,superscriptaddress]{revtex4-1}

\usepackage{amssymb}
\usepackage{amsmath}
\usepackage{graphicx}
\usepackage{amsbsy}
\usepackage{color}
\usepackage{bm}

\newcommand{\bq}{\begin{eqnarray}}
\newcommand{\eq}{\end{eqnarray}}
\newcommand{\bqn}{\begin{eqnarray*}}
\newcommand{\eqn}{\end{eqnarray*}}
\newcommand{\rr}{\mathbf{r}}

\begin{document}
\title{Gas-liquid coexistence for the bosons square-well fluid and the
$\mbox{}^4$He binodal anomaly}

\author{Riccardo Fantoni}
\email{rfantoni@ts.infn.it}
\affiliation{Dipartimento di Scienze Molecolari e Nanosistemi,
  Universit\`a Ca' Foscari Venezia, Calle Larga S. Marta DD2137,
  I-30123 Venezia, Italy} 


\date{\today}


\begin{abstract}
The binodal of a boson square-well fluid is determined as a function
of the particle mass through the newly devised quantum Gibbs ensemble
Monte Carlo algorithm [R. Fantoni and S. Moroni, {\sl to be
published}]. In the infinite mass limit we recover the classical
result. As the particle mass decreases the gas-liquid critical point
moves at lower temperatures. We explicitely study the case of a
quantum delocalization de Boer parameter close to the one of
$\mbox{}^4$He. For comparison we also determine the gas-liquid
coexistence curve of $\mbox{}^4$He for which we are able to observe
the binodal anomaly below the $\lambda$-transition temperature.  
\end{abstract}

\maketitle
\label{sec:introduction}

Soon after Feynman rewriting of quantum mechanics and quantum
statistical physics in terms of the path integral
\cite{Feynman1948,FeynmanFIP} it was realized that the new
mathematical object could be used as a powerful numerical instrument. 

The statistical physics community soon realized that a path integral
could be calculated using the Monte Carlo method \cite{Ceperley1995}.

Consider a fluid of $N$ bosons at a given absolute temperature $T=
1/k_B\beta$ with $k_B$ Boltzmann constant. Let the system of particles
have a Hamiltonian $\hat{H}=-\lambda\sum_{i=1}^N 
\bm{\nabla}_i^2+\sum_{i<j}\phi(|\rr_i-\rr_j|)$ symmetric under
particle exchange, with $\lambda=\hbar^2/2m$, $m$ the mass of the
particles, and $\phi(|\rr_i-\rr_j|)$ the pair-potential of interaction
between particle $i$ at $\rr_i$ and particle $j$ at
$\rr_j$. The many-particles system will have spatial
configurations $\{R\}$, with $R\equiv (\rr_1,\ldots,\rr_N)$ the
coordinates of the $N$ particles. The partition function of the fluid
can be calculated \cite{Ceperley1995} as a sum over the $N!$ possible
particles permutations, ${\cal P}$, of a path integral over
closed many-particles paths $X\equiv(R_0,\ldots,R_P)$ in the
imaginary time interval $\tau\in[0,\beta=P\epsilon]$, discretized into
$P$ intervals of equal length $\epsilon$, the time-step, with
$R_P={\cal P}R_0$ the $\beta$-periodic boundary condition.

More recently a grand canonical ensemble algorithm has been devised by
Massimo Boninsegni et al. \cite{Boninsegni2006a,*Boninsegni2006b} for
the path integral Monte Carlo method. This paved the way to the
development of a quantum Gibbs ensemble Monte Carlo algorithm (QGEMC)
to study the gas-liquid coexistence of a generic boson fluid
\cite{Fantoni2014}. This algorithm is the quantum analogue of
Athanassios Panagiotopoulos
\cite{Panagiotopoulos87,*Panagiotopoulos88,*Smit89a,*Smit89b,*Frenkel-Smit}
method which has now been successfully used for several decades to study
first order phase transitions in classical fluids
\cite{Panagiotopoulos92,*Sciortino2009,*Fantoni2013}. However, like
simulations in the 
grand-canonical ensemble, the method does rely on a reasonable number
of successful particle insertions to achieve compositional
equilibrium. As a consequence, the Gibbs ensemble Monte Carlo method
cannot be used to study equilibria involving very dense phases.
Unlike previous extensions of Gibbs ensemble Monte Carlo to include
quantum effects (some \cite{Schneider1995,*Nielaba1996} only consider
fluids with internal quantum states; others
\cite{Wang1997,*Georgescu2013,*Kowalczyk2013} successfully exploit the
path integral Monte Carlo isomorphism between quantum particles and
classical ring 
polymers, but lack the structure of particle exchanges which underlies
Bose or Fermi statistics), the QGEMC scheme is viable even for systems
with strong quantum delocalization in the degenerate regime of temperature. 
Details of the QGEMC algorithm will be presented elsewhere
\cite{Fantoni2014}.   

In this communication we will apply the QGEMC method to the fluid of
square well bosons in three spatial dimensions as an extension of the
work of Vega et al. \cite{Vega1992,*Liu2005} on the classical
fluid. The de Boer quantum delocalization parameter
$\Lambda=\hbar/\sigma(m{\cal E})^{1/2}$, with ${\cal E}$ and $\sigma$
measures of the energy and length scale of the potential energy, can
be used to estimate the quantum mechanical effects on the
thermodynamic properties of nearly classical liquids
\cite{Young1980}. We will consider square well 
fluids with two values of the particle mass $m$: $\Lambda=1/\sqrt{50}$,
close but different from zero, and $\Lambda=1/\sqrt{5}$. In the first
case we compare our result with the one of Vega and in the second case
with the one of $\mbox{}^4$He which we consider in our
second application. When studying the binodal of $\mbox{}^4$He in
three spatial dimensions we are able to reproduce the binodal anomaly
appearing below the $\lambda$-point where the liquid branch of the
coexistence curve shows a re-entrant behavior.

\label{sec:algorithm}

In our implementation of the QGEMC \cite{Fantoni2014} algorithm we
choose the primitive approximation to the path integral action 
discussed in Ref. \cite{Ceperley1995}. The simulation is
performed in two boxes (representing the two coexisting phases) of
varying volumes $V_1$ and $V_2=V-V_1$ and numbers of particles
$N_1=V_1\rho_1$ and $N_2=V_2\rho_2=N-N_1$ with $V$ and $N=V\rho$
constants. The Gibbs equilibrium conditions of pressures and chemical
potentials equality between the two boxes is enforced by allowing
changes in the volumes of 
the two boxes (the {\sl volume move}, $q=5$) and by allowing exchanges
of particles between the two boxes (the {\sl open-insert move}, $q=1$,
plus the complementary {\sl close-remove move}, $q=2$, plus the {\sl
  advance-recede move}, $q=3$) while at the same time sampling the
closed paths configuration space (the {\sl swap move}, $q=4$, plus the
{\sl displace move}, $q=6$, plus the {\sl wiggle move}, $q=7$). We
thus have a menu of seven, $q=1,2,\ldots,7$, different Monte Carlo
moves where a single random attempt of any one of them with
probability $G_q=g_q/\sum_{q=1}^7g_q$ constitutes a Monte Carlo step.

We denote with ${\cal V}$ the maximum displacement of
$\ln(V_1/V_2)$ in the volume move, with ${\cal L}^{(p)}$ the
maximum particle displacement in box $p=1,2$ in the displacement move,
and with ${\cal M}_q<P$ the maximum number of time slices involved in the
$q\neq 5,6$ move. In order to fulfill detailed balance we must choose
${\cal M}_1={\cal M}_2$. 
 
Letting the system evolve at a given absolute temperature $T$ from a
given initial state (for example we shall take $\rho_1=\rho_2=\rho$) we
measure the densities of the two coexisting phases, $\rho_1<\rho$ and
$\rho_2>\rho$, which soon approach the coexistence equilibrium values.

\label{sec:gas-liquid}

\label{sec:sw-model}
First we study a system of bosons in three dimensions interacting with
a square well pair-potential
\bq
\phi(r)&=&\left\{\begin{array}{ll}
+\infty & r<\sigma\\
-{\cal A}      & \sigma\le r <\sigma(1+\Delta)\\
0       & \sigma(1+\Delta)\le r
\end{array}\right.
\eq
which, for example, can be used as an effective potential
for cold atoms \cite{Pethik-Smith} with a scattering length
$a=\sigma(1+\Delta)[1-\tan(\sigma\Delta\sqrt{{\cal
A}/2\lambda})/\sigma(1+\Delta)\sqrt{{\cal A}/2\lambda}]$.
We choose ${\cal A}>0$ as the unit of energies and $\sigma$ as the unit
of lengths. We then introduce a reduced temperature $T^*=k_BT/{\cal
  A}$ and a reduced density $\rho^*=\rho\sigma^3$. When the mass of
the boson is very big, i.e. $\lambda^*=\lambda/({\cal A}\sigma^2) \ll
1$ we are in the classical limit. The classical fluid has been studied
originally by Vega {\sl et al.} 
\cite{Vega1992} who found that the critical point of the gas-liquid
coexistence moves at lower temperatures and higher densities as
$\Delta$ gets smaller. 
The quantum mechanical effects on the thermodynamic properties of
nearly classical liquids can be estimated by the de Boer quantum
delocalization parameter $\Lambda=\sqrt{2\lambda^*}$.  

\label{sec:results}

During the subcritical temperature runs we register the densities of
the gas, $\rho_g$, and of 
the liquid, $\rho_l (>\rho_g)$, phase (box). When the densities of the
two boxes are too close one another we may observe curves crossing which
implies that the two boxes exchange identity. It is then necessary
the computation of a density probability distribution function,
created using the densities of both boxes. When we are at temperatures
sufficiently below the critical point, this distribution appears to be
bimodal, i.e. it has two peaks approximated by 
Gaussians. In some representative cases we checked that the peaks of
the bimodal so calculated occur at the same densities as the peaks of
the bimodal obtained from the single density distribution of the worm
algorithm after a careful tuning of the chemical potential
\cite{Wilding1995}. 
 
We study the model with $\Delta=0.5$ near their classical 
limit $\lambda^*=1/100$ ($\Lambda\approx 0.14, a^*=a/\sigma\approx
1.44$) and at an intermediate case $\lambda^*=1/10$ ($\Lambda\approx
0.45, a^*\approx 0.58$). We choose $N=50$, $\rho^*=0.3$, ${\cal
  L}^{(p)}=V_p^{1/3}/10$, ${\cal V}=1/10$, we take all ${\cal M}_q$
equal, adjusted so as to have the acceptance ratios of the {\sl wiggle}
move close to $50$\%, $g_1=g_2=g_3=g_4=g_7=1$, $g_5=0.0001$, and
$g_6=0.1$. Moreover we choose the relative weight of the Z and G
sectors of our extended worm algorithm, $C$ \cite{Boninsegni2006a},  
so as to have the Z-sector acceptance ratios close to $50$\%. We  
started from an initial configuration where we have an equal number 
of particles in boxes of equal volumes at a total density $\rho^*=0.3$.

All our runs were made of $10^5$ blocks of $10^5$ MC steps with 
properties measurements every $10^2$ steps 
\footnote{Our QGEMC code took $\approx 90$ seconds of CPU time for one
  million steps of a system of size $N=50, P=10, {\cal M}_q=5$ calculating
  properties every $100$ steps, on an IBM iDataPlex DX360M3 Cluster
  (2.40GHz). The algorithm scales as $N^2$, due to the potential
  energy calculation, and as $P$, due to the volume move.}. The time
needed to reach the equilibrium coexistence increases with $P$ and in
general with a lowering of the temperature. 

\begin{figure}[htbp]
\begin{center}
\includegraphics[width=8cm]{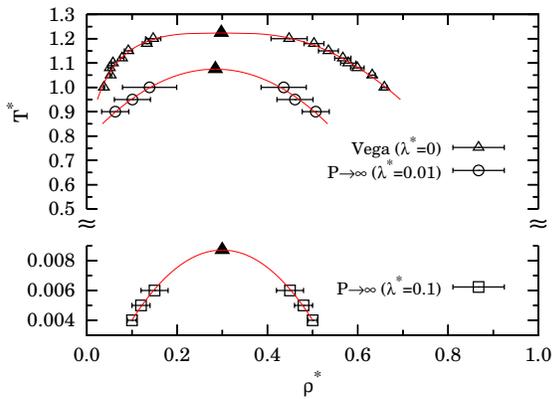}
\end{center}  
\caption{(color online) Binodal for the square well fluid in three
  dimensions. Shown 
  are the classical results of Vega et al. \cite{Vega1992} at
  $\lambda^*=0$ and our results in the $P\to\infty$ limit for
  $\lambda^*=1/100,1/10$. In the simulations we used 
  $N=50$ and for the extrapolation to the zero time-step limit up to
  $P=20$ for $\lambda^*=1/100$ and $P=500$ for $\lambda^*=1/10$. The
  curves extrapolating to the critical point are obtained as described
  in the text. The filled triangles are the expected critical points.} 
\label{fig:sw}
\end{figure}

\begin{figure}[htbp]
\begin{center}
\includegraphics[width=8cm]{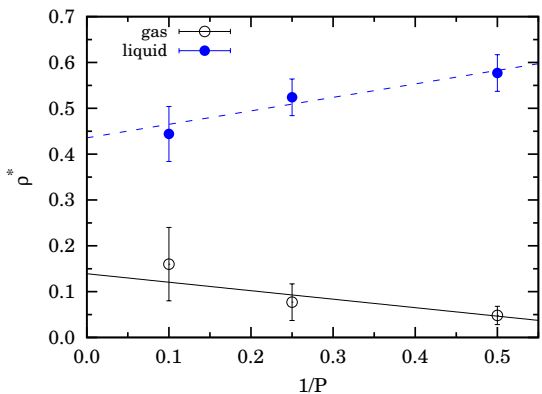}
\end{center}  
\caption{(color online) Linear fit to the zero time-step limit
  $P\to\infty$ for $T^*=1$ and $\lambda^*=1/100$.} 
\label{fig:fit}
\end{figure}

If we choose $\lambda^*=1/100$ and $P=2$, ${\cal M}_q=1$ (in this  
case the advance-recede move cannot occur) we find that our algorithm
gives results close to the ones of Vega \cite{Vega1992} obtained
with the classical statistical mechanics ($\lambda^*=0$) algorithm of
Panagiotopoulos \cite{Panagiotopoulos87,Smit89a,Smit89b}
\footnote{Note that there is no difference between our algorithm in
the limit $P=2, {\cal M}_q=1$, and $\lambda^*\to 0$ and the one of
Panagiotopoulos \cite{Panagiotopoulos87,Smit89a,Smit89b}.}. 
As we diminish the time-step  
$\epsilon^*=1/PT^*$ at a given temperature we can extrapolate to the
zero time-step limit $P\to \infty$ as shown in 
Fig. \ref{fig:fit}. We thus obtain the fully quantum statistical
mechanics result for the binodal shown in Fig. \ref{fig:sw} which
turns out to exist for $T^*\lesssim 1$. This shows
that the critical point due to the effect of the quantum statistics
moves at lower temperatures. For the studied temperatures the
superfluid fraction \cite{Pollock1987} of the system was always
negligible as in the systems studied in
Ref. \cite{Wang1997,*Georgescu2013,*Kowalczyk2013} like Neon
($\Lambda\approx 0.095$) and molecular Hydrogen ($\Lambda\approx
0.276$).   

In order to extrapolate the binodal to the critical point we used the
law of ``rectilinear diameters'', $\rho_l+\rho_g=2\rho_c+a|T-T_c|$,
and the Fisher expansion \cite{Fisher1966},
$\rho_l-\rho_g=b|T-T_c|^{\beta_1}(|T-T_c|+c)^{\beta_0-\beta_1}$, with
$\beta_1=1/2$ and $\beta_0=0.3265$, and $a,b,c$ fitting parameters
with $c=0$ for $\lambda=0$ and $c\neq 0$ for $\lambda\neq 0$.  


Upon increasing $\lambda^*$ to $1/10$ the binodal now appears at
$T^*\lesssim 0.008$ where we had a non negligible superfluid fraction
\cite{Pollock1987} ($\rho_s/\rho\approx 0.32(2)$ at $T^*=0.006$ on the
liquid branch). As a consequence it proves necessary to use bigger
$P$ in the extrapolation to the zero time-step limit. Notice also that
at lower temperature it is necessary to run longer simulations due to
the longer paths and equilibration times. We generally expect that
increasing $\lambda^*$ the gas-liquid critical temperature decreases
and the normal-super fluid critical temperature increases. So the
window of temperature for the normal liquid tends to close. 

\label{sec:4He-model}

Our second study is on $\mbox{}^4$He, for which $\lambda^*=6.0596$. We
now take $1$\r{A} as unit of lengths and $k_B$K 
as unit of energies. In this case $\sigma\approx 2.5$\r{A}, ${\cal
  E}\approx 10.9$K, and $\Lambda\approx 0.42$. A situation comparable
to the square well case with $\lambda^*=1/10$. We use $N=128$ and the
Aziz HFDHE2 pair-potential \cite{Aziz1979}
\bq
\phi(r)&=&\left\{\begin{array}{ll}
\epsilon\phi^*(x) & r<r_\text{cut}\\
0                 & r\geq r_\text{cut}
\end{array}\right.,\\
\phi^*(x)&=&A\exp(-\alpha x)-
\left(\frac{C_6}{x^6}+\frac{C_8}{x^8}+\frac{C_{10}}{x^{10}}\right)F(x),\\
F(x)&=&\left\{\begin{array}{ll}
\exp[-(D/x-1)^2] & x<D\\
1                & x\geq D
\end{array}\right.,
\eq
where $x=r/r_m$, $r_m=2.9673$, $\epsilon/k_B=10.8$, $A=0.5448504$,
$\alpha=13.353384$, $C_6=1.3732412$, $C_8=0.4253785$,
$C_{10}=0.178100$, $D=1.241314$, and $r_\text{cut}=6$\r{A} (here we
explicitly checked that during the 
simulation the conditions $V_p^{1/3}>2r_\text{cut}$ for $p=1,2$ are
always satisfied). In this case it proves convenient to choose
$\rho^*=0.01$, ${\cal L}^{(p)}=V_p^{1/3}/10$, ${\cal
V}=1/10$, $g_1=g_2=g_3=g_4=g_7=1$,
$g_5=0.0001$, and $g_6=0.1$. As for the SW case we observe a
decrease of the width of the coexistence curve $\rho_l-\rho_g$ as the
number of time slices increases. We thus work at a small (fixed)
time-step $\epsilon^*=0.002$ about $1/1000$ of the superfluid
transition temperature as advised in Ref. \cite{Ceperley1995} to be
necessary when studying Helium with the primitive approximation for 
the action.  

\begin{figure}[htbp]
\begin{center}
\includegraphics[width=8cm]{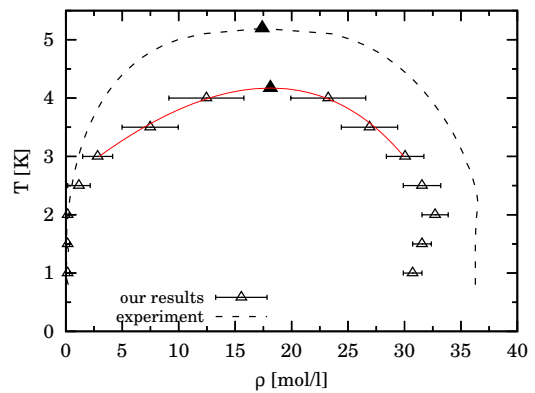}
\end{center}  
\caption{(color online) Binodal for the $\mbox{}^4$He of Aziz
  \cite{Aziz1979} in three dimensions. In our simulations we used
  $N=128$, $r^*_\text{cut}=6$, and a time-step $\epsilon^*=0.002$. The
  continuous (red) curve extrapolating to the 
  critical point are obtained as described in the text. The filled
  triangle is the estimated critical point. The experimental results
  from Ref. \cite{McCarty1980} are also shown as a dashed
  curve.}    
\label{fig:he4}
\end{figure}

The results for the binodal are shown in Fig. \ref{fig:he4}. The
experimental critical point is at $T_c=5.25$K and $\rho_c=17.3$mol/l
\cite{McCarty1973,*McCarty1980}. Factors explaining the discrepancy
with experiment could be the size error or the choice of the
pair-potential. Choosing bigger sizes $N$ it is possible to increase
$r_\text{cut}$ and this shifts the simulated critical temperature to
higher values. For the three dimensional $\mbox{}^4$He we expect to
have the superfluid below a $\lambda$-temperature $T^*_\lambda=2.193(6)$
\cite{Boninsegni2006b}, so our results again show that our method works
well even in the presence of a non negligible superfluid
fraction. Moreover as shown by the points at the two lowest
temperatures we are observing the expected \cite{Stein1998} binodal
anomaly below the $\lambda$-point.  

%

\label{sec:conclusions}

In conclusion we determined the gas-liquid binodal of a square well
fluid of bosons as a function of the particle mass and of
$\mbox{}^4$He, in three spatial dimensions, from first principles. The
critical point of the square well fluid moves to lower temperatures as
the mass of the particles decreases, or as the de Boer parameter
increases, while the critical density stays approximately constant.  

Our results for $\mbox{}^4$He compare well with the experimental
critical density even if a lower critical temperature is observed in
the simulation. We expect this to be due mainly to a finite size
effect unavoidable in the simulation. Nonetheless we are able to
determine the binodal anomaly \cite{Stein1998} occurring below the
$\lambda$-transition temperature. The anomaly 
that we observe in the simulation appears to be more accentuated than
in the experiment and the liquid branch of the binodal falls at
slightly lower densities.  

Even if our QGEMC method is more efficient at high temperatures it is
able to detect the liquid phase at low temperatures even below the
superfluid transition temperature. The new numerical method is
extremely simple to use and unlike current methods does not need the
matching of free energies calculated separately for each phase or the
simulation of large systems containing both phases and their
interface. 

\appendix
\label{app:1}

R.F. would like to acknowledge the use of the PLX computational
facility of CINECA through the ISCRA grant. We are grateful to
Michael Ellis Fisher for correspondence and helpful comments.
\bibliography{qgemc}
\end{document}